\documentclass[superscriptaddress,twocolumn,amsmath,amssymb]{revtex4-1}

\usepackage[colorlinks=true,bookmarks=false,citecolor=blue,urlcolor=blue]{hyperref} 

\usepackage{amsmath,amssymb}
\usepackage{graphicx}
\usepackage{verbatim}
\usepackage{color}
\usepackage[english]{babel}
\usepackage{bm}
\usepackage{natbib}
\usepackage[symbol]{footmisc}
\usepackage{textcomp}
\newcommand{\red}{\textcolor{red}}
\newcommand{\blue}{\textcolor{blue}}

\usepackage{setspace}

\begin{document}

\preprint{APS/123-QED}

\title{All-optical nanoscale heating and thermometry with resonant dielectric nanoparticles for photoinduced tumor treatment}
\thanks{A footnote to the article title}%

\author{George P. Zograf}
\affiliation{Department of Nanophotonics and Metamaterials, ITMO University, St. Petersburg, 197101, Russia}

\author{Alexander S. Timin}
\altaffiliation{Research School of Chemical and Biomedical Engineering, National Research Tomsk Polytechnic University, Tomsk, 634050, Russia}

\author{Albert P. Muslimov}
\altaffiliation{Peter the Great St. Petersburg Polytechnic University, Saint-Petersburg, 195251, Russia}
\affiliation{First I. P. Pavlov State Medical University of St. Petersburg, Saint-Petersburg, 197022, Russia}

\author{Ivan I. Shishkin}
\affiliation{Department of Nanophotonics and Metamaterials, ITMO University, St. Petersburg, 197101, Russia}

\author{Alexandre Nomin\'e}
\affiliation{Department of Nanophotonics and Metamaterials, ITMO University, St. Petersburg, 197101, Russia}

\author{Jaafar Ghanbaja}
\affiliation{Institute Jean Lamour, CNRS, Universit\'e de Lorraine, Nancy, 54011, France}

\author{Pintu Ghosh}
\author{Qiang Li}
\affiliation{State Key Laboratory of Modern Optical Instrumentation, College of Optical Science and Engineering, Zhejiang University, Hangzhou 310027, China}

\author{Mikhail V. Zyuzin}
\email{mikhail.zyuzin@metalab.ifmo.ru}

\author{Sergey V. Makarov}
\email{s.makarov@metalab.ifmo.ru}
\affiliation{Department of Nanophotonics and Metamaterials, ITMO University, St. Petersburg, 197101, Russia}

\date{\today}

\begin{abstract}
All-dielectric nanophotonics becomes a versatile tool for various optical applications, including nanothermometry and optical heating. Its general concept includes excitation of Mie resonances in nonplasmonic nanoparticles. However, the potential of resonant dielectric nanoparticles in drug delivery applications still have not been fully realized. Here, optically resonant dielectric iron oxide nanoparticles ($\alpha$-Fe$_2$O$_3$ NPs) are employed for remote rupture of microcontainers used as drug delivery platform. It is theoretically and experimentally demonstrated, that $\alpha$-Fe$_2$O$_3$ NPs has several advantages in light-to-heat energy conversion comparing to previously used materials, such as noble metals and silicon, due to the broader spectral range of efficient optical heating, and in enhancement of thermally sensitive Raman signal. The $\alpha$-Fe$_2$O$_3$ NPs embedded into the wall of universal drug carriers, polymer capsules, are used to experimentally determine the local temperature of the capsule rupture upon laser irradiation (170$^o$C). As a proof of principle, we successfully show the delivery and remote release of anticancer drug vincristine upon lowered laser irradiation (4.0$\times$10$^4$~W/cm$^2$) using polymer capsules modified with the $\alpha$-Fe$_2$O$_3$ NPs. The biological tests were performed on two primary cell types: (i) carcinoma cells, as an example of malignant tumor, and (ii) human stem cells, as a model of healthy cells. 
The developed delivery system consisting of polymer capsules modified with the dielectric nanoparticles provides multifunctional platform for remote drug release and temperature detection.
\end{abstract}

\maketitle


\section{Introduction}

Optically resonant dielectric nanoparticles (NPs) allowing for light localization at subwavelength scale have attracted a lot of attention~\cite{staude2017metamaterial, kuznetsov2016optically} during last years due to their excellent performance in a number of advanced optical applications. Namely, the resonant NPs made of silicon~\cite{evlyukhin2012demonstration, tittl2018imaging}, germanium~\cite{grinblat2016enhanced}, III-V semiconductors~\cite{liu2016resonantly, cambiasso2017bridging}, or multicomponent materials~\cite{timpu2016second, noskov2018non, tiguntseva2018light} demonstrate outstanding optical characteristics owing to Mie resonances excited by incident laser field. 
Moreover, because of strong field enhancement \textit{inside} the nonplasmonic (e.g., silicon) NPs, it is possible to both efficiently heat them by laser irradiation and measure local temperature via thermally sensitive Raman  response~\cite{zograf2017resonant}, which is also enhanced by several orders of magnitude~\cite{dmitriev2016resonant}. 
With the resonant silicon NPs, the ``all-in-one'' concept was applied for simultaneous optical heating, sensing, and thermometry of small amount of protein molecules~\cite{milichko2018metal}. 
However, the full potential of resonant nonplasmonic NPs with \textit{both} temperature and micrometer spatial control for more advanced bioapplications has yet to be realized.

Recent trends have focused on incoprorating different NPs within nano- and microcarriers to obtain a composite platform with improved properties for various biomedical applications \cite{kantner2018laterally}. Among available delivery systems, polymer capsules are currently being demonstrated as unique carriers for safe and efficient delivery of biologically active compounds such as gene materials~\cite{ott2015light}, macromolecules~\cite{zyuzin2017comprehensive}, proteins~\cite{karamitros2013preserving}, drugs~\cite{timin2018multifunctional}. In the previous works, various plasmonic NPs were employed as nanoheaters to reduce laser intensity threshold for the capsules opening, while their additional functionalization by thermally sensitive nanomaterials was performed for local nanothermometry~\cite{yashchenok2015optical}.

In this work, we incorporate optically resonant dielectric nanoparticles into polymer capsules for remote drug release upon laser irradiation with simultaneous nanothermometry.
In order to achieve high optical field enhancement in the nanoparticles and their efficient heating, we employ biocompatible iron oxide ($\alpha$-Fe$_2$O$_3$) NPs~\cite{laurent2008magnetic} instead of previously used dielectric ones~\cite{tittl2018imaging, timpu2016second,tiguntseva2018light}, owing to their high refractive index ($n \approx$2.7--3.2) and moderate losses ($k \approx$0.02--0.2) in red and near-infrared (NIR) ranges~\cite{akl2004optical}. This phase of iron oxide has very pronounced and easily recognizable multipeak Raman response in broad range of wavenumbers (200--1400~cm$^{-1}$~\cite{maslar2000situ}), making it much more convenient for temperature-induced peak shift registration as compared to the silicon possessing only one Raman-active phonon mode~\cite{milichko2018metal}. The use of such NPs allows for direct measurement of the capsules rupturing temperature with the accuracy no worse than 40~K, corresponding to a characteristic Raman linewidth. Also, the NPs decrease the rupture threshold intensity down to 4.0$\times$10$^4$~W/cm$^2$. Finally, the developed microcapsules modified with $\alpha$-Fe$_2$O$_3$ NPs are used to deliver antitumor drug vincristine (VCR) into two primary cell types: (i) carcinoma cells (CCs) as an example of malignant tumor, and (ii) human mesenchymal stem cells (hMSCs) as a model of healthy cells. Our findings show that the optically resonant $\alpha$-Fe$_2$O$_3$ NPs have high potential to be a versatile platform in order to fabricate multifunctional drug carriers possessing tumor-specific drug release.

\section{Results and discussion}

\begin{figure*}[ht!]
  \centering
  \includegraphics[width=0.8\textwidth]{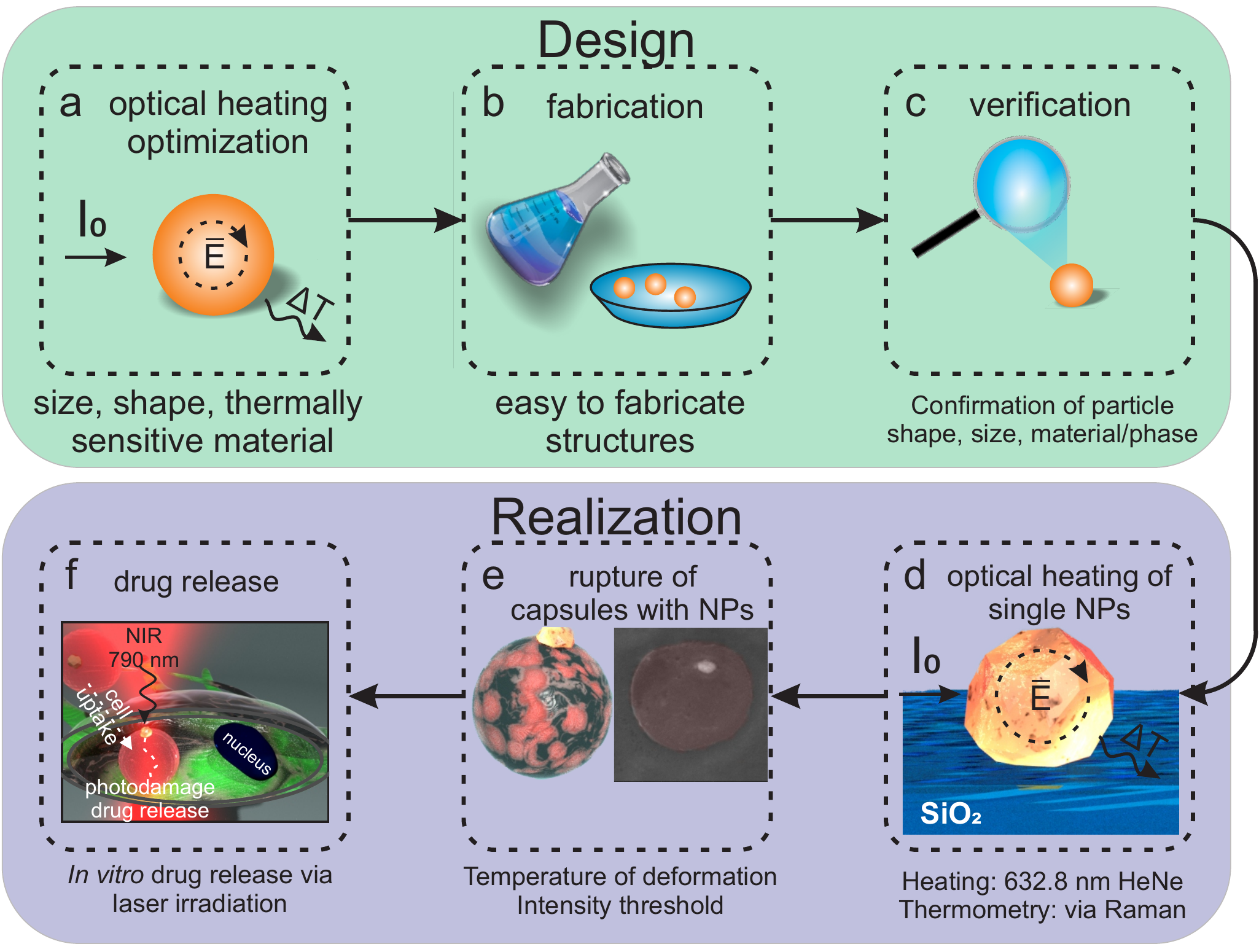}\\
  \centering
  \caption{\textbf{Schematic representation of iron oxide NPs implementation steps in this work.} (a) Simulation of optimal optical heating parameters. (b) Synthesis of NPs with the required physicochemical properties. (c) Characterization of obtained NPs. (d) Study of NPs optical properties. (e) Immobilization of NPs into the shell of polymer capsules for drug delivery applications. (f) Probing the remote drug activation inside cells and nanothermometry.} \label{outline}
\end{figure*}

The main concept of this study is depicted in Fig.~\ref{outline}, where we present the theoretical and experimental stages of designing a multifunctional drug carrier that simultaneously allows for external drug release triggering and local temperature detection. 
As shown in Fig.~\ref{outline}a, the used biocompatible $\alpha$-Fe$_2$O$_3$ NPs should possess high enough refractive index and moderate losses, as well as demonstrate thermally sensitive Raman lines. Also, to predict the optimal NP size for efficient optimal heating, the electromagnetic simulations should be performed (Fig.~\ref{outline}a). After the theoretical calculations, synthesized (Fig.~\ref{outline}b) and characterized (Fig.~\ref{outline}c) NPs can be experimentally tested for optical heating and Raman thermometry (Fig.~\ref{outline}d). The fully characterized iron oxide NPs can be then used to modify the wall of polymer capsules, which are widely used to deliver cargo into cells (Fig.~\ref{outline}e). The efficient optical heating of the NPs allows for remote low-intensity release of cargo inside cells (Fig.~\ref{outline}f). In the following sections, we discuss each step in more details.

\begin{figure*}[ht!]
  \centering
  \includegraphics[width=0.98\textwidth]{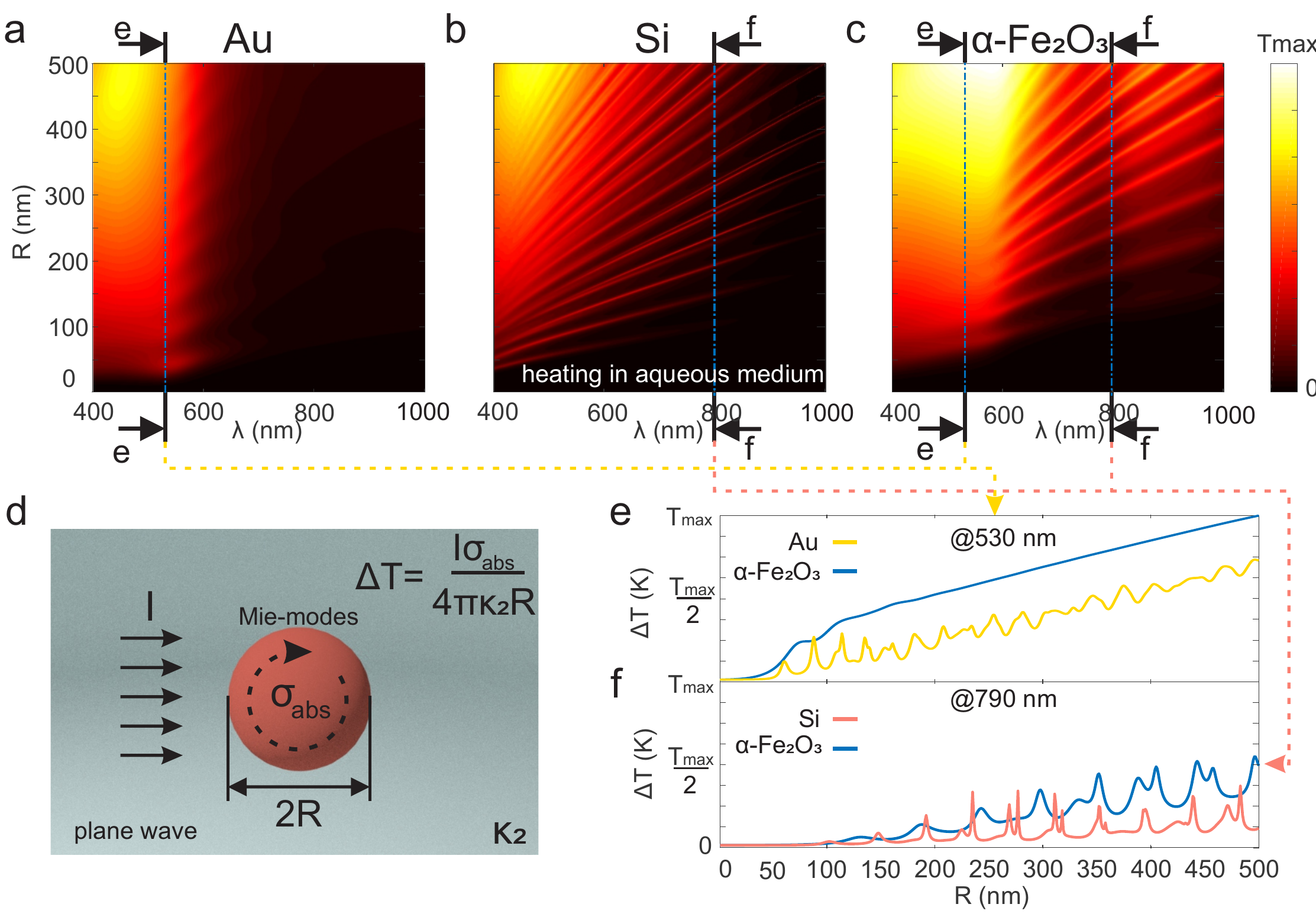}\\
  \centering
  \caption{\textbf{Theoretical modeling of NPs optical heating.} Calculated maps for optical heating of (a) Au, (b) Si and (c) $\alpha$-Fe$_2$O$_3$ spherical NPs in water using Eq.~(\ref{eq:46}--\ref{eq:temperature}). The colorbar is temperature increase in arbitrary units. (d) Schematic of optical heating theoretical model. Incident plane-wave excites resonant optical modes of a sphere (Mie-theory) in homogeneous surrounding medium with thermal conductivity \(\kappa_2\). 1D plots for the dependence of $\Delta T$ on NP radius $R$ extracted from Fig.~\ref{Map}a,b for fixed excitation wavelengths: \(\lambda\) = 530~nm for Au (e) and  \(\lambda\) = 790~nm for Si (f) compared with heating of $\alpha$-Fe$_2$O$_3$ NPs taken from Fig.~\ref{Map}c.} \label{Map}
\end{figure*}

\textbf{Theoretical background.} General model for heating of a spherical nanoparticle comprised from an arbitrary material (metals or dielectrics) by continuous laser illumination in a homogeneous medium was developed elsewhere~\cite{skirtach2005role, baffou2010nanoscale,zograf2017resonant}. We combined it with a simple analytical solution for light absorption by the nanoparticle from the Mie theory~\cite{bohren2008absorption}. The Mie-theory deals with a single spherical NP in homogeneous medium, yielding general understanding which optical modes can be excited and what is their optical response. For instance, scattering ($\sigma_{sca}$), extinction ($\sigma_{ext}$) and absorption ($\sigma_{abs}$) cross-sections can be derived from following equations:
\begin{equation}\label{eq:46}
\sigma_{sca} = \frac{W_{sca}}{I_i} = \frac{2\pi}{k^2} \sum^{\infty}_{n=1} (2n+1) (|a_n|^2+|b_n|^2)
\end{equation}
\begin{equation}\label{eq:47}
\sigma_{ext} = \frac{W_{ext}}{I_i} = \frac{2\pi}{k^2} \sum^{\infty}_{n=1} (2n+1) Re(a_n+b_n)
\end{equation}
\begin{equation}\label{eq:4700}
\sigma_{abs} = \sigma_{ext} - \sigma_{sca},
\end{equation}
where $a_{n}$ and $b_{n}$ are the Mie-coefficients for electric and magnetic types of modes, respectively. As a result, the solution of a thermal diffusion equation in a steady-state regime with a heat source defined by the light intensity ($I$) and the NP absorption cross-section ($\sigma_{abs}$) gives the following equation for temperature increase inside the NP~\cite{baffou2010nanoscale} 
:
\begin{equation}\label{eq:temperature}
\delta T_{NP}=\frac{I\sigma_{abs}}{4\pi \kappa_2 R},
\end{equation}
where $\kappa_2$ is the thermal conductivity of the surrounding medium, which is significantly smaller than that of the NP; and $R$ is the NP radius. The details of derivation of this formula and calculations of the absorption cross-section one can find in \textit{Supporting Information}. It is worth noting, that this consideration is only suitable for structures with thermal conductivity of several orders higher than that of surrounding homogeneous media.

Since the concept of NP-induced enhancement and localization of optical heating~\cite{baffou2013thermo} is attractive for the most of bio- or chemical applications, it is vital to study their resonant properties in water (n~=~1.33), because such studies are taken in aqueous medium. The theoretical calculations on various NPs optical heating in water are summarized in Fig.~\ref{Map}. Based on Eq.(\ref{eq:46}--\ref{eq:temperature}) we carried out a comparative analysis for optical heating in the visible and NIR ranges of three spherical nanoparticles: gold~\cite{johnson1972optical}, silicon~\cite{aspnes1983dielectric}, iron oxide (Fe$_2$O$_3$)~\cite{querry1985optical}, while their thermal constants are taken from Ref.~\cite{speight2005lange}.

It is clearly seen from Fig.~\ref{Map}a that gold nanospheres are efficient light-to-heat converters at their localized surface plasmon resonance wavelength (around $\lambda\sim$~500--600~nm) at all sizes, as reported previously~\cite{baffou2013thermo}. On the other hand, a silicon nanosphere, as shown in Fig.~\ref{Map}b, supports variety of Mie modes (electric and magnetic ones~\cite{kuznetsov2016optically}), which allow to use c-Si nanoparticles as light-to-heat converters even in the red region of visible light spectra~\cite{zograf2017resonant}. However, their performance worsens at the wavelengths larger than $\lambda\approx$~800~nm. In its turn, $\alpha$-Fe$_2$O$_3$ NPs combine benefits of both gold and silicon by providing the broadband absorption band in the visible at $\lambda~<$~600~nm wavelengths and support Mie-modes in NIR as well (Fig.~\ref{Map}c). Additionally, a comparison of performances between $\alpha$-Fe$_2$O$_3$, c-Si and Au nanospheres is presented in Fig.~\ref{Map}c,d, where one can clearly see that the  $\alpha$-Fe$_2$O$_3$ NP outperforms the Au NPs even at the plasmon resonance wavelength in water $\lambda\approx$~530~nm. Moreover, the $\alpha$-Fe$_2$O$_3$ NP absorbs light better than Si NPs at all wavelengths of interest ($\lambda$~=~400~--~1000~nm). The origin of such outstanding performance of Fe$_2$O$_3$ is based on the optimized combination of  real and imaginary parts of its dielectric function. Indeed, radiative and non-radiative losses of any resonators have to be balanced to achieve the highest optical heating~\cite{zograf2017resonant}.

\begin{figure*}[ht!]
  \centering
  \includegraphics[width=0.9\textwidth]{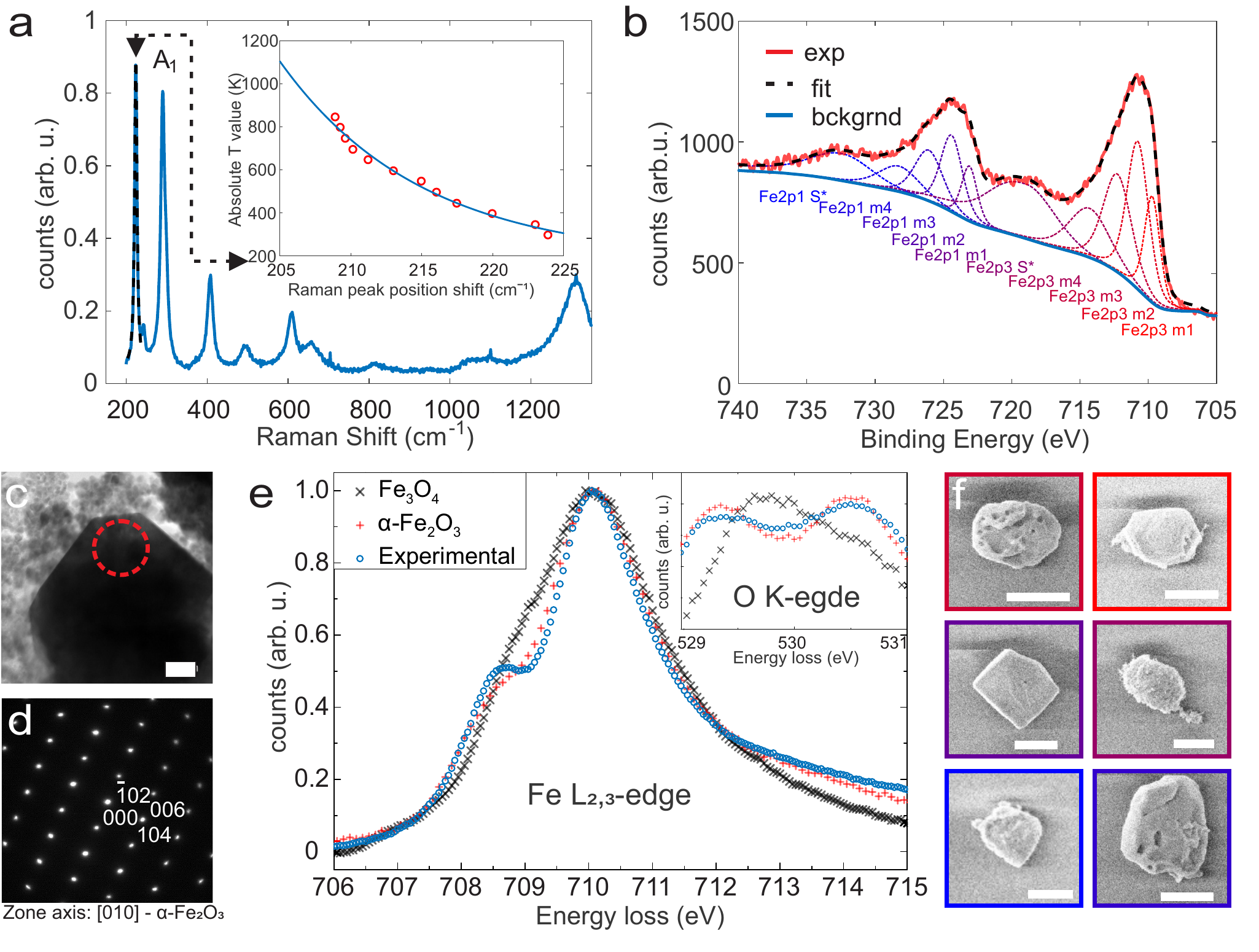}\\
  \caption{\textbf{$\alpha$-Fe$_2$O$_3$ NPs characterization.} (a) Stokes Raman scattering from fabricated $\alpha$-Fe$_2$O$_3$ NPs. Inset illustrates evolution of A\(_1\) band position with increase in temperature (adopted from~\cite{rout2009temperature}) (b) X-ray photoelectron spectroscopy of fabricated structures. Captions under curves correspond to specific response of Fe atoms. (c) TEM image of the iron oxide NPs. Scale bar is 50~nm. Dashed region corresponds to the area of consideration for (d) electron diffraction patterns. (e) Electron Energy Loss Spectroscopy (EELS) for the same NPs of $\alpha$-Fe$_2$O$_3$. (f) Typical shapes, sizes and morphologies of fabricated $\alpha$-Fe$_2$O$_3$ NPs carried out by means of scanning electron microscopy (scale bar is 400~nm).} \label{characterization}
\end{figure*}

Beside high efficiency of light-to-heat conversion, it is crucial to obtain a pronounced Raman signal for local optical temperature measurements. This concept of far-field optical temperature detection at nanoscale is based on thermo-sensitive Raman response due to anharmonic effects in light scattering on optical phonons~\cite{balkanski1983anharmonic}. Fig.~\ref{characterization}a shows that $\alpha$-Fe$_2$O$_3$ has very intensive and rich Raman spectrum with clearly distinguishable Stokes lines.

As a result, in this part we revealed that $\alpha$-Fe$_2$O$_3$ is one of the most prospective materials for all-optical heating and nanothermometry. Namely, it fulfills the following requirements: (i) relatively simple and cost-efficient fabrication; (ii) active optical phonons for thermo-sensitive optical response (Raman scattering); (iii) high-refractive index in order to support Mie-resonances in nanostructures in the visible and NIR for enhanced heating and reduced signal collection (dwell) time.

\textbf{Nanoparticles fabrication and structural characterization.} $\alpha$-Fe$_2$O$_3$ NPs were prepared via solvothermal method~\cite{mitra2007synthesis}. The characterization by scanning electron microscopy (SEM) found NPs with diameter ($D$) between 300 and 800~nm (Fig.~\red{S3}). That is further confirmed by transmission electron microscopy (TEM), which also shows a second population of NPs with diameters in the range of 5-20~nm (Fig.~\red{S5}). Despite the larger $\alpha$-Fe$_2$O$_3$ NPs do not possess a perfect spherical shape, the aspect ratio remains sufficiently low to maintain the quasi-spherical approach reasonable for theoretical calculations, since the factor of facets for resonant dielectric NPs is almost negligible as shown in previous works~\cite{timpu2016second, tiguntseva2018light}.

X-ray photoelectron spectroscopy (XPS) results confirm that NPs are composed of iron oxide (Fig~\ref{characterization}b). However, the peaks in the obtained spectra correspond to both +II and +III oxidation states (\textit{i.e.} \(\alpha\)-Fe$_2$O$_3$ and Fe$_3$O$_4$, respectively). The challenge to determine composition and crystallogaphic strucutre of NPs arises due to the fact that two phases share either the same stoichiometry ($\alpha$-Fe$_2$O$_3$ hematite, and $\gamma$-Fe$_2$O$_3$ maghemite) or structural similarities ($\gamma$-Fe$_2$O$_3$ maghemite and $\gamma$-Fe$_3$O$_4$ magnetite)\cite{ChenPhysRevB.79.104103}. However, \(\alpha\)-Fe$_2$O$_3$ and Fe$_3$O$_4$ can be additionally differentiated by electron energy loss spectroscopy (EELS)~\cite{ewels_sikora_serin_ewels_lajaunie_2016}. In the case of the large NPs shown in Fig.~\ref{characterization}c, the experimental EELS spectrum shows the typical features of \(\alpha\)-Fe$_2$O$_3$, which are a double peak in the O K-edge and a shoulder in the Fe L$_\textsc{2,3}$ edge Fig.~\ref{characterization}e. Knowing the stoichimetry, selected area is used to determine the structure of the same NP. The diffraction pattern shown in Fig.~\ref{characterization}d revealed, that such NPs have high crystallinity and interplanar distances match very well with the diffraction pattern of $\alpha$-Fe$_2$O$_3$ in [010] zone axis (Fig.~\red{S6}). Characterization of smaller particles was carried out by following the same procedure. Unlike in the case of larger NPs, EELS spectra as well as diffraction pattern correspond to Fe$_3$O$_4$ magnetite (Fig.~\red{S7}). This explains the detection of both oxidation state on the XPS spectra.
The presence of both oxides is not necessarily an issue for this study, since Fe$_3$O$_4$ shows only limited optical heating for particle radius below 50~nm (Fig.~\red{S1}). Moreover, since there is a correlation between size and crystallographic phase, the smaller NPs can be easily removed during the centrifugation process (e.g., during capsule synthesis).

As it has been mentioned, $\alpha$-Fe$_2$O$_3$ NPs can serve not only as a nanoscale heater, but also simultaneously as multi-purpose nanothermometer~\cite{maslar2000situ} with high melting point ($T\approx$1850~K~\cite{zhao2007iron}) as compared to gold ($T\approx$1250~K~\cite{dick2002size}) and silicon ($T\approx$1600~K~\cite{hull1999properties}). Thus, Fig.~\ref{characterization}a depicts not only phase and bond fingerprint of fabricated structure, but also can serve as temperature sensor, because of thermally dependent Raman response. For instance, A$_1$ band (vibration of Fe-O bond~\cite{chamritski2005infrared,maslar2000situ}) scattering demonstrate spectral blue-shift, as it is shown in the inset~\cite{rout2009temperature}. For higher precision of the temperature measurements, we average the temperature between 225~cm$^{-1}$, 290~cm$^{-1}$ and 400~cm$^{-1}$ Raman Stokes lines according to previous works~\cite{maslar2000situ}. 

\begin{figure*}[ht!]
  \centering
  \includegraphics[width=0.9\textwidth]{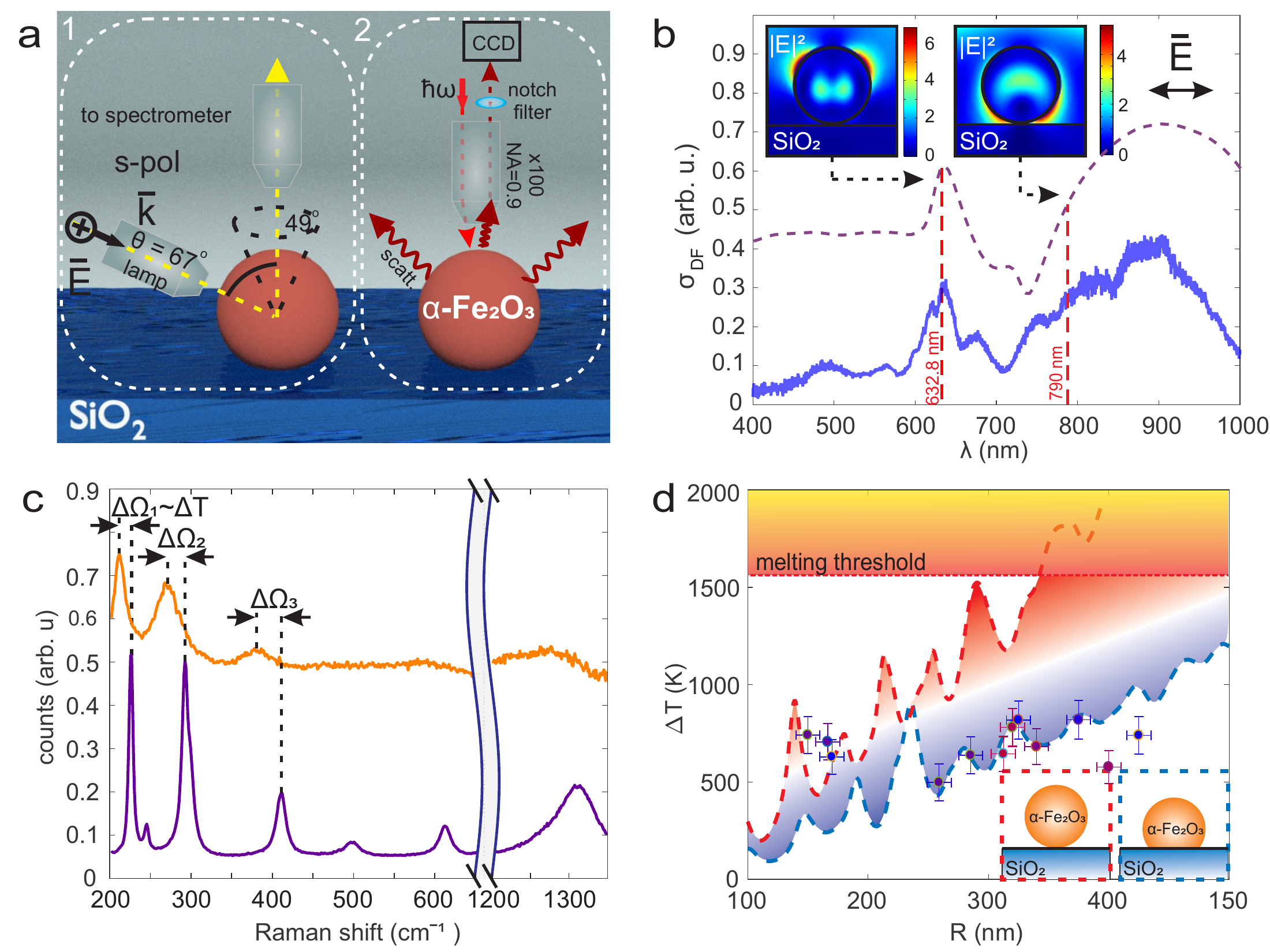}\\
  \caption{\textbf{Optical heating of $\alpha$-Fe$_2$O$_3$ NPs} (a) Schematic of (1) dark-field (DF) elastic scattering spectroscopy and (2) Stokes-Raman spectroscopy experimental setups. (b) Dark-field scattering spectra of a single $\alpha$-Fe$_2$O$_3$ NP (radius $R~=~$170~nm): experimental (solid blue) and numerically calculated (dashed purple). Insets depict calculated electric field distribution at corresponding wavelengths (632.8~nm and 790~nm) corresponding to the wavelengths of laser sources for the optical heating experiments. (c) Stokes range of Raman scattering from single the $\alpha$-Fe$_2$O$_3$ NP from (b) at two intensities \blue{0.6}~mW/$\mu$m$^2$ (`cold regime', purple line) and \blue{2.4}~mW/$\mu$m$^2$ (`hot regime', orange line). (d) Experimental (dots) and theoretical (dashed lines) dependencies of optical heating of a single $\alpha$-Fe$_2$O$_3$ NP upon irradiation by CW-laser with intensity \blue{2.4}~mW/$\mu$m$^2$ at wavelength 632.8~nm. Blue dashed and red dashed lines correspond to theoretical calculations with different thermal contact areas between the nanoparticle and SiO\(_2\) substrate, as shown in the insets, where a nanosphere is dipped by R/2.5 and 0 nm into the substrate, respectively).
}\label{Raman}
\end{figure*}

\textbf{Optical characterization of resonant $\alpha$-Fe$_2$O$_3$ NPs.}
In order to study resonant optical properties of fabricated single $\alpha$-Fe$_2$O$_3$ NPs, dark-field (DF) spectroscopy technique was applied (Fig.~\ref{Raman}a,b, for more details see \emph{Supporting Informaion}). It is worth noticing, that DF spectra serves not only as a tool for characterization of the far-field optical properties and description of the excited optical modes in NPs, but also for the verification of the NPs sizes obtained by SEM or TEM. In order to carry out the `optical size' verification, the full-wave numerical simulations in the commercial software COMSOL Multiphysics was employed. As shown in Fig.~\ref{Raman}b, the numerical and experimental results are in a good qualitative agreement, even though the fabricated NPs are not perfectly spherical according to Fig.~\ref{characterization}c,f. The modeling also reveals the local optical field enhancement up to 2 times within NP and up to 6 times outside NP (see insets in Fig.~\ref{Raman}b). 

The DF spectrum shows that the chosen NP (340 nm diameter particle) provides resonant elastic scattering around the emission wavelength of He-Ne laser ($\lambda$~=~632.8~nm) used for Raman experiments in this work. This results in the local field enhancement and NP`s heating observed as a spectral shift of characteristic Raman-active modes (Fig.~\ref{Raman}c,~\red{S10}). As it has been mentioned, the spectral shift of the Raman Stokes line between 'cold' and 'hot' regimes can be back-converted into temperature increase. By term 'cold' Raman scattering regime either bulk material response, or low pump intensity response from single iron oxide nanoparticle is considered. Whereas the 'hot' regime stands for enhanced absorption regime with high ($>$ 2~mW/$\mu$m$^2$) pump intensity, which results in spectral blue-shift of the Raman response. It is clearly seen, that for such high heating temperatures, only number of Raman lines remain pronounced (i.e. initial 225~cm$^{-1}$), unlike 600 and 1320~cm$^{-1}$ ones.

Fig.~\ref{Raman}d depicts both theoretical and experimental optical heating results for NP with radius $R\approx$~170~nm at 632.8~nm wavelength with intensity $I_0$~=~2.4~mW/$\mu$m$^2$. According to SEM images, $\alpha$-Fe$_2$O$_3$ NPs are not perfectly spherical, thus the thermal contact area determination remains only one fitting parameter for our numerical calculations. In this regard, blue dashed line corresponds to dipped by diameter over 5 ($R$/2.5) nanometers into SiO$_2$ substrate and red dashed line is the case of a slight physical contact (20~nm into substrate). The experimental values are mostly located within the region bordered by two calculated lines, however, larger particles are in little less match with the theoretical predictions. We believe that this might happen due to shape inhomogeneity. Indeed, any roughness reduces quality factor of the dielectric microresonator~\cite{borselli2005beyond}, whereas the heating efficiency is a linear function of quality factor~\cite{zograf2017resonant}.

\begin{figure*}[ht!]
  \centering
  \includegraphics[width=0.98\textwidth]{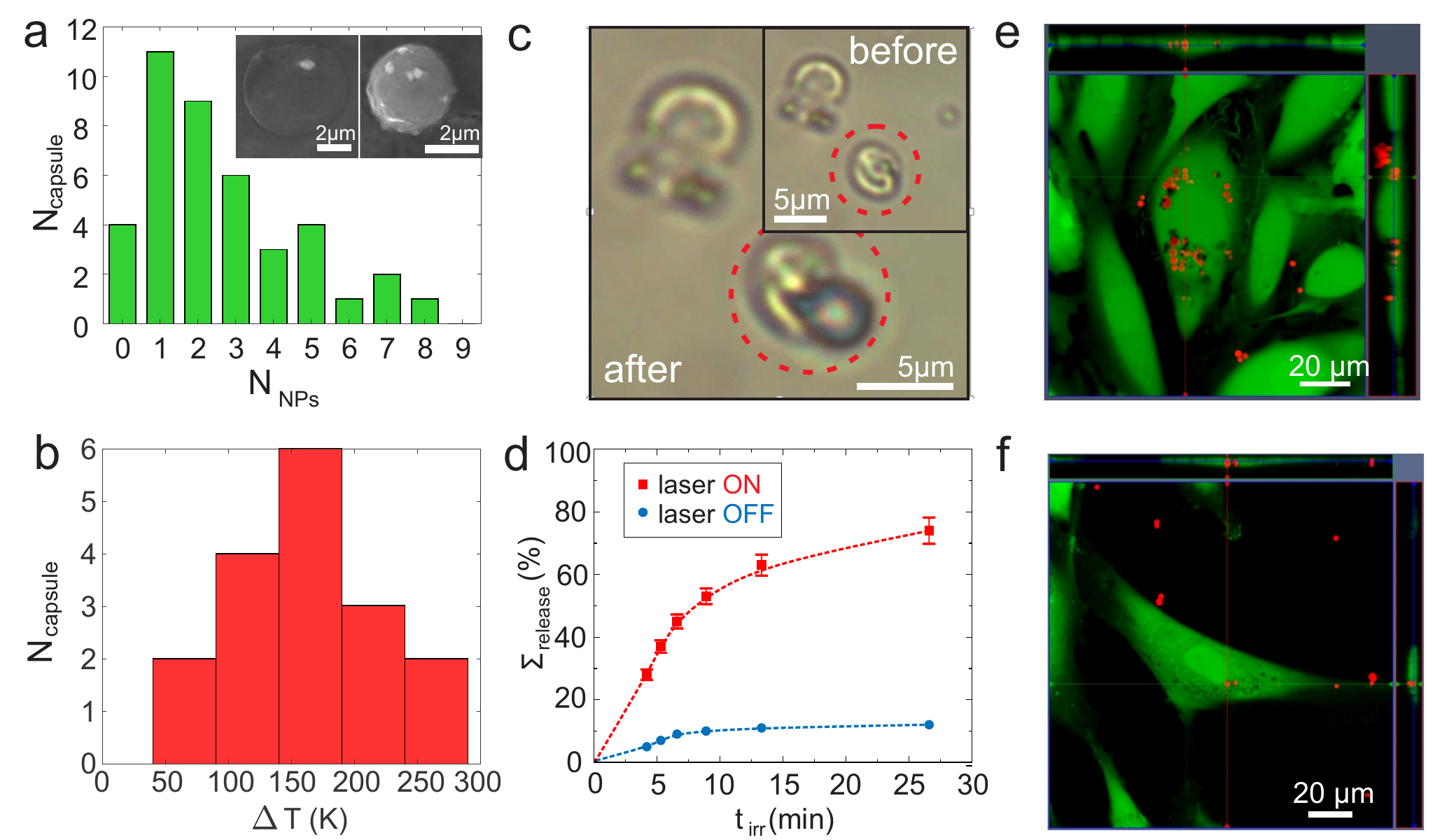}\\
  \caption{\textbf{$\alpha$-Fe$_2$O$_3$ NPs in polymer capsule.} (a) Number of $\alpha$-Fe$_2$O$_3$ NPs embedded into capsules walls according to SEM images. Inset: SEM images of fabricated capsule with embedded $\alpha$-Fe$_2$O$_3$ NPs into walls. Scale bar is 2~\(\mu m\). (b) Optical thermal destroy temperature distribution according to Raman experiments. Experiments were performed on empty capsules. (c) Bright field images of capsule before (inset) and after laser irradiation. (d) Percentage of released antitumor drug VCR from  polymer capsules irradiated during different time periods (red curve) and without irradiation (blue curve). (e) Orthogonal view from different planes (x/y, x/z or y/z) of the confocal microscope images used to analyze the particle uptake. Example corresponds to hMSCs with internalized capsules. Cells were staied with calcein AM (shown in green) and capsules were labelled with AF647 (shown in red). (f) Orthogonal view from different planes (x/y, x/z or y/z) of the CLSM images used to analyze the capsules uptake. Example corresponds to CCs with internalized capsules. Cells were staied with calcein AM (shown in green) and capsules were labelled with AF647 (shown in red). }
  \label{capsule}
\end{figure*}

\textbf{Polymer capsules with resonant $\alpha$-Fe$_2$O$_3$ NPs.}
After proving the ability of both heating and thermometry by single iron oxide NPs  (Fig.~\ref{outline}d), we implemented them into drug carriers, so-called polymer capsules, to probe remote drug activation in vitro and simultaneous nanothermometry (Fig.\ref{outline}e). It is worth noting that various carriers composed of different materials like silica~\cite{liu2015gold}, liposomes~\cite{luo2016doxorubicin}, as well as rare earth~\cite{bagheri2016lanthanide} and plasmonic nanoparticles~\cite{goodman2016understanding} were employed for external triggering to release bioactive molecules upon laser irradiation. However, polymer capsules have an obvious advantages prior reported delivery systems, such as increased drug loading capacity, protective features of capsule wall, low cytotoxicity, high stability in the biological fluids, and efficient uptake by various cell types~\cite{timin2017multi, zyuzin2017comprehensive}. Polymer capsules modified with $\alpha$-Fe$_2$O$_3$ NPs were prepared with Layer-by-Layer method~\cite{hussain2016catalysis,zyuzin2017comprehensive}. The size range of synthesized capsules was around 4.38 - 5.21~$\mu$m, as determined by SEM and confocal laser scanning microscopy (CLSM) analysis (Fig.\ref{capsule}a(inset) and Fig.~\red{S8}a,b). It can be seen that $\alpha$-Fe$_2$O$_3$ NPs were not aggregated after attachment to the capsules wall Fig.~\red{S8}.

CLSM was used to verify the loading with dextran conjugated with AlexaFluor 647 (AF647) into the cavity of the capsules (Fig.~\red{S8}b). SEM images demonstrated successful dissolution of CaCO$_3$ core-template, since the polymer capsules were in collapsed form under vacuum (Fig.~\red{S8}a). The attachment of $\alpha$-Fe$_2$O$_3$ NPs onto the capsules wall (spots onto the capsules wall) can be clearly seen in SEM, dark field (DF) and CLSM images highlighted with the white arrows Fig.~\red{S8}a-c. The statistical data for the number of attached $\alpha$-Fe$_2$O$_3$ NPs onto the capsules wall is shown in Fig.~\ref{capsule}a. 
Successful loading of commercially available antitumor drug VCR into the capsules cavity could be detected with indirect method via measuring absorption spectra of the drug. According to the measured calibration curve Fig.~\red{S4} and measured supernatant after the formation of the CaCO$_3$ core-particles, amount of encapsulated VCR was 0.65$\times$10$^{-12}$~g/cap. It was assumed that after the core dissolution, the amount of VCR, which diffused through the capsule wall, is negligibly low.

\begin{table*}[ht!]
\centering
\caption{Comparison of the proposed resonant $\alpha$-Fe$_2$O$_3$ NPs with previously published design for enhanced optical heating and simultaneous nanothermometry.}
\begin{tabular}{|c|c|c|c|c|}
\hline
Approach                                                     & \begin{tabular}[c]{@{}c@{}}Capsule damage \\ intensity,\\ wavelength\end{tabular} & \begin{tabular}[c]{@{}c@{}} Thermometry \\ method \end{tabular}                                                       & \begin{tabular}[c]{@{}c@{}} Temperature \\ resolution \end{tabular} & Ref       \\ \hline
\begin{tabular}[c]{@{}c@{}}Gold NPs \\ with CNT\end{tabular} & \begin{tabular}[c]{@{}c@{}}9.6$\times$10$^4$ W cm$^{-2}$\\ $\lambda$=532~nm\end{tabular}                        & \begin{tabular}[c]{@{}c@{}}Raman:\\ Stokes/Anti-Stokes ratio\end{tabular} & 5-40 K                 & \cite{yashchenok2015optical} \\ \hline
$\alpha$-Fe$_2$O$_3$ NPs                                          & \begin{tabular}[c]{@{}c@{}}4.0$\pm$1.0$\times$10$^4$ W cm$^{-2}$ \\ $\lambda$=633~nm\end{tabular}                          & \begin{tabular}[c]{@{}c@{}}Raman:\\ Stokes spectral shift\end{tabular}    & 40 K                   & This work \\ \hline
\end{tabular}
\end{table*}

In order to determine the temperature of the capsules rupture upon laser illumination, we carried out their Raman nanothermometry at various intensities of incident laser light. Employing the developed Raman nanothermometry with the resonant $\alpha$-Fe$_2$O$_3$ NPs incorporated into the capsule walls, we measured Raman spectra for different capsules up to the signal disappearance as an indicator of the capsule's destruction. According to the measurements from a set of capsules with incorporated NPs placed as a droplet between two cover glasses, the mean temperature of their rupture is around 170~$^o$C (Fig.~\ref{capsule}b). The fact of capsule damage was confirmed by bright-field optical microscopy (Fig.~\ref{capsule}c). The average intensity of He-Ne laser radiation ($\lambda$=632.8~nm) required to achieve this threshold temperature is around 4$\times$10$^4$~W/cm$^2$. The value of temperature is comparable with that reported elsewhere~\cite{yashchenok2015optical}, while the applied intensity is more than 2 times lower, because of high efficiency of light-to-heat conversion of $\alpha$-Fe$_2$O$_3$ NPs, as shown in Fig.~\ref{Map}. The comparison of main characteristics of the proposed method with published before is given in Table~1.
To confirm that permeability of polymer capsules changes after laser irradiation and delivered cargo can be successfully released, polymer capsules modified with $\alpha$-Fe$_2$O$_3$ NPs were loaded with commercially available antitumor drug VCR and the defined area containing capsules was irradiated with near infrared laser for different periods of time. As it can be seen in Fig.~\ref{capsule}d, with the increasing time of irradiation within the area containing capsules, the amount of released drug increases. However, the capsules that were not irradiated also showed slight drug release, which can be explained by leakage of the cargo through the porous capsules walls.

The obtained capsule heating temperature (170~$^o$C ) is sufficient to change the permeability of polymer capsules, since the glass transition temperature of PAH is around 85~$^o$C~
and PSS is around 152~$^o$C~\cite{niidome2000enormous}. The photodecomposition of polymer capsules can lead to the release of loaded molecules into the surroundings. In comparison to the previous work~\cite{yashchenok2015optical}, to achieve temperature 170~$^o$C less laser power density is required (4.0$\pm$1.0$\times$10$^4$~W cm$^{-2}$), which is in favor for the biomedical laser application, since such laser power densities do not induce phototoxicity of cells. Indeed, no phototoxicity was observed even at intensity 3.8$\times$10$^5$~W cm$^{-2}$ as shown in ref.~\cite{carregal2012nir}. Moreover, according to Eq.~S30, such high for biological objects temperature decays by more than 10 times on a scale of several microns around the locally heated NP, being safe for the irradiated living cell containing an internalized capsule modified with the \(\alpha\)-Fe$_2$O$_3$ NPs.

\textbf{Cell uptake and association of polymer capsules modified with $\alpha$-Fe$_2$O$_3$ NPs.}
It is worth mentioning that cell experiments were performed on relevant cell models: (i) clear cell renal cell carcinoma cells (CCs), and (ii) human mesenchymal stem cells (hMSCs), which were derived from patients who signed a voluntary consent. Numerous studies suggest that cell lines poorly represent the diversity, heterogeneity and drug-resistant tumors occurring in patients~\cite{zhou2009tumour}. The derivation and short--term culture of primary cells from solid tumors have, thus, gained significant importance in personalized cancer therapy. CCs are sensitive to VCR at high dosages but not sensitive to the concentrations of this drug used in clinics~\cite{hartmann1999chemotherapy}. hMSCs were used in this study as a model of healthy cells.

In order to verify whether micrometric polymer capsules modified with $\alpha$-Fe$_2$O$_3$ NPs can be internalized with CCs and hMSCs, CLSM measurements were performed. For this, after incubation with different amount of polymer capsules loaded with AF647, cells were stained with calcein AM and the co-localization of the red fluorescently labeled capsules within the cellular compartments was visualized with $z$-stack images as shown in Fig.~\ref{capsule}e,f. An indicator for intracellular localization of capsules was red fluorescence signal coming from AF647 labeled capsules surrounded with green fluorescence signal coming from stained living cell with calcein AM. Fig.~\red{S16} shows 3D reconstruction of cells incubated with capsules. These results are in agreement with previous studies, where it was shown that even micrometric sized non-targeted polymer capsules can be efficiently internalized with stem cells~\cite{timin2018multifunctional} and carcinoma cells~\cite{zyuzin2017comprehensive}. Note that multiple endocytotic pathways are responsible for micrometric capsules internalization~\cite{kastl2013multiple}. 

In order to quantify the association of AF647 labeled capsules at different capsule concentrations with CCs and hMSCs, fluorescence flow cytometry was used. Here, the amount of red fluorescence originating from each analyzed cell was derived and quantified to reveal the relative amount of capsules associated with cells. It is worth to mention that apart from autofluorescence cells do not contribute to the red fluorescence signal Fig.~\red{S14}. According to the obtained data, it is clearly seen that amount of capsules associated with cells increases with the number of added capsules per cell. This is valid for the both cell lines. However, it is also seen that the capsules were associated with CCs in a higher rate than with hMSCs (Fig.~\red{S15}). These results are in agreement with previous works~\cite{huang2007multifunctional}. As it has been discussed above, multiple internalization pathways are involved in micrometric sized capsules internalization. Internalization of capsules depends on the metabolism of the cells, which defines their ability to internalize the capsules. Metabolism of carcinoma cells is known to be enhanced~\cite{zhang2015polymer} and distorted, which explains the higher uptake rate of capsules by CCs than by hMSCs.

\begin{figure*}[ht!]
  \centering
  \includegraphics[width=0.98\textwidth]{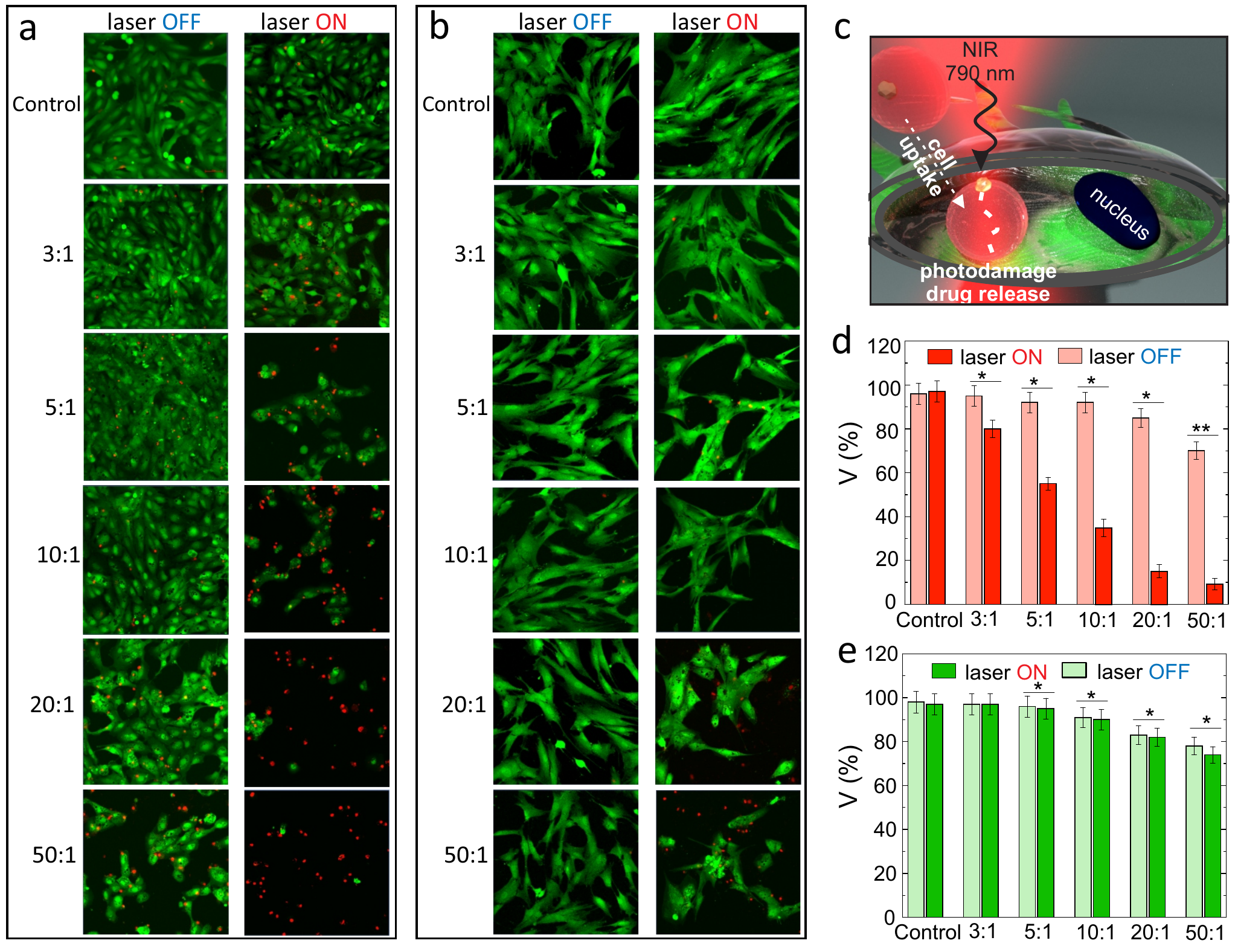}\\
  \caption{\textbf{$\alpha$-Fe$_2$O$_3$ NPs in capsules for remote drug delivery upon laser irradiation.} (a) CLSM images of CCs incubated with different amounts of polymer capsules modified with$\alpha$-Fe$_2$O$_3$ NPs and loaded with antitumor drug VCR for 24 h. CCs were not irradiated with NIR laser (right column) and CCs were irradiated with laser (left column). Living cells were stained with calcein AM (green) and dead cells with propidium iodide (red). (b) CLSM images of hMSCs incubated with different amounts of polymer capsules modified with $\alpha$-Fe$_2$O$_3$ NPs and loaded with antitumor drug VCR for 24 h. hMSCs were not irradiated with NIR laser (right column) and hMSCs were irradiated with laser (left column). Living cells were stained with calcein AM (green) and dead cells with propidium iodide (red). The scale bar for (a,b) is 50~$\mu$m. (c) Schematic illustration of experimental setup: cross-section of a cell with internalized capsule inside. Capsule is irradiated with NIR 790 nm laser. (d) CCs viability after incubation with polymer capsules modified with $\alpha$-Fe$_2$O$_3$ NPs and loaded with antitumor drug VCR for 24~h before (red) and after (pink) irradiation with NIR laser. (e) hMSCs viability after incubation with polymer capsules modified with $\alpha$-Fe$_2$O$_3$ NPs and loaded with antitumor drug VCR for 24~h before (green) and after (bright green) irradiation with NIR laser. The results are presented as the average value $\pm$ standard deviation (* represents p $<$~0.05 and ** represents p $<$~0.005). 
}\label{release}
\end{figure*}

\textbf{In vitro laser-triggered release of antitumor drug Vincristine (VCR).} 
To demonstrate remote photoinduced activation of drug, $\alpha$-Fe$_2$O$_3$ NPs modified polymer capsules loaded with VCR were added to the CCs and hMSCs at different amounts. Next day, cells were washed 3 times with cell culture medium in order to remove residual capsules, and then irradiated with NIR ($\lambda$=790~nm) in the defined area (Fig.~\ref{release}c,\red{S11}). NIR laser irradiation was used for the experiments with cells, since this wavelength lies in the `tissue window', and it is optimal for penetration of light into tissue~\cite{hainfeld2014infrared}. Similar wavelengths are often used in the biology and medicine for therapy~\cite{he2018crucial}. As shown in Fig.~\ref{release}, $\alpha$-Fe$_2$O$_3$ NPs efficiently convert light into heat also at $\lambda$=790~nm, being enough for the capsules rupture in a multipulse mode owing to high repetition rate (80~MHz), as also confirmed by calculations in Fig.~\red{S2}. During the NIR laser irradiation at such `CW-like' regime with an average power of 250~mW, laser beam randomly hits the capsules inside the scanning area. After irradiation of capsules with NIR light, $\alpha$-Fe$_2$O$_3$ NPs exhibit strong absorption supported by Mie-modes, which results in conversion of light-to-heat energy followed by the NPs temperature increase and, subsequently, in the rupture of capsules. In other words, the interaction between the laser beam and $\alpha$-Fe$_2$O$_3$ NPs embedded into the capsules wall leads to the thermally induced deformation/disintegration of the polymer capsules with the change in their permeability, as observed in Fig.~\ref{capsule}c. 

In the control experiments, the cells without capsules were irradiated with NIR laser at the same conditions, as well as the cells without and with capsules (added at different amounts) were not irradiated with NIR laser. After irradiation, LIVE/DEAD viability assay was performed, where the cells alive were stained in green and nuclei of dead cells were stained in red. According to the obtained data, the used power density of laser irradiation was not toxic for CCs and hMSCs (Fig.~\ref{release}a,b). As expected, laser irradiation of CCs with internalized/associated capsules resulted in higher toxicity, due to the release of antitumor drug VCR in intra- and extracellular environment. As it can be seen in Fig.~\ref{release}d, this toxicity of CCs is dependent on the amount of added capsules per cell. The possible explanation for this can be that the probability to hit the capsule added at amount of 50 capsules per cell is higher and, thus, higher amounts of drugs were released from the capsules associated with cells, which resulted in higher toxicity. This assumption is confirmed by the performed release study where different amount of VCR loaded capsules modified with iron oxide nanoparticles were irradiated under the same conditions but without cells Fig.~\red{S13}. The obtained results demonstrate that higher amount of added capsules results in a higher rate of released drugs after laser irradiation keeping the irradiation time constant.  Interestingly, hMSCs showed significantly higher survival rate compared to CCs at the same added amount of capsules. The reason for this could be the higher uptake of tumor cells, what results in higher amount of antitumor drug inside cells. Moreover, hMSCs are resistant to the number of antitumor drugs, what can be a reason of higher survival rate of hMSCs compared to CCs~\cite{bellagamba2016human}.

Capsules added to the both cell models, which were not irradiated with NIR laser, were slightly toxic at higher concentrations (86\% for 20 caps/cell, 70\% for 50 caps/cell for CCs and 84\% for 20 caps/cell, 78\% for 50 caps/cell for hMSCs). This can be explained by leakage of drugs through the capsules wall during the internalization process. Indeed, during the internalization process~\cite{kantner2018laterally} mechanical deformation can occur, which can result in the leakage of drug. According to the recent study, mechanical forces can vary between cell types, and it was shown that capsules internalized within carcinoma cells are exposed to higher mechanical forces during the internalization process in carcinoma cells than in other tested cell types resulting in a higher rate of capsule degradation~\cite{chen2016analysing}, which can lead to a higher drug release rate inside cells. Moreover, it has been shown that capsules added at amount more than 20 capsules per cell are toxic~\cite{lepik2016mesenchymal} to the cells resulting in reduced cell density in Fig.\ref{release}b. Thereby, the revealed data confirm that $\alpha$-Fe$_2$O$_3$ NPs can be effectively used as heat mediators for remote activation of drugs inside cells.

\section{Conclusion}

To summarize, we have proposed novel ``all-in-one'' concept for controllable drug release in living cells based on nonplasmonic optically resonant nanoparticles which allowed for lowering the drug release irradiation threshold and local thermometry. It has been proven both theoretically and experimentally that $\alpha$-Fe$_2$O$_3$ NPs have several advantages in terms of light-to-heat energy conversion comparing to conventional materials, such as noble metals and silicon: (i) $\alpha$-Fe$_2$O$_3$ NPs demonstrate broader electromagnetic radiation absorption in the visible and near-IR ranges; (ii) require lesser irradiation dose compared to noble metals and dielectric NPs and do not require strict fulfilling of Mie-resonant conditions to achieve similar heating temperatures; (iii) support number of thermally-sensitive Raman lines in the range from (200~cm$^{-1}$) to (1400~cm$^{-1}$), which is more convenient for nanothemrometry and might be useful for tracing additional molecular events in real time. Moreover, simplicity of $\alpha$-Fe$_2$O$_3$ NPs synthesis and its biocompatibility make this material suitable for biomedical applications. 
Taken together, we believe that such iron oxide based nanoscale structures will serve as a versatile platform, that combines unique optical properties with laser responsive effect for the drug release, proving complete suppression the growth of the primary tumor cells.

\section{acknowledgement}

Part of this work related to the synthesis of nanomaterials was supported by the
Russian Science Foundation, grant No. 17-19-01637 (M.V.Z.). M.V.Z. also thanks the President’s Scholarship SP-1576.2018.4. The biological experiments concerning cells interaction with capsules were supported by the Russian Science Foundation, No. 17-73-10023 (A.S.T.). Visualization of real-time capsules rupture was done with support of National Key Research and Development Program of China (2017YFE0100200). G.Z. thanks Dr. Soslan Khubezhov for XPS experiments, Dr. Mikhail Zhukov and Filipp Komissarenko for SEM investigations and Dr. Mihail Petrov for valuable discussion.

\bibliography{main.bib}

\end{document}